\documentclass[british,aps,prb,twocolumn]{revtex4-1}
\usepackage[T1]{fontenc}
\usepackage[latin9]{inputenc}
\usepackage{color}
\usepackage{amsmath}
\usepackage{graphicx}
\usepackage{babel}
\usepackage{lineno}
\usepackage{cleveref}

\begin{document}

\title{{\normalsize{}Protected ground states in short chains of coupled spins in circuit quantum electrodynamics}}

\author{{\normalsize{}Adam Callison\textsuperscript{1}, Eytan Grosfeld\textsuperscript{2}
and Eran Ginossar\textsuperscript{1}}}

\affiliation{\textsuperscript{1}Advanced Technology Institute and Department
of Physics, University of Surrey, Guildford GU2 7XH, United Kingdom}

\affiliation{\textsuperscript{2}Department of Physics, Ben-Gurion University
of the Negev, Be\textquoteright er-Sheva 84105, Israel}
\begin{abstract}
The two degenerate ground states of the anisotropic Heisenberg (XY) spin model of a chain of qubits (pseudo-spins) can encode quantum information, but their degree of protection against local perturbations is known to be only partial. We examine the properties of the system in the presence of non-local spin-spin interactions, possibly emerging from the quantum electrodynamics of the device. We find a phase distinct from the XY phase admitting two ground states which are highly protected against all local field perturbations, persisting across a range of parameters. In the context of the XY chain we discuss how the coupling between two ground states can be used to observe signatures of topological edge states in a small controlled chain of superconducting transmon qubits. 
\end{abstract}

\maketitle

There is much interest in the use of a topological ground-state degeneracy in quantum systems to realise a quantum bit where quantum information can be encoded \cite{dennis2002topmem}. In principle, such states can  overcome significant obstacles to robust quantum computing since it is theorised that the topological nature of the encoding, essentially very non-local, would be protecting against sources of locally-acting noise \cite{nayak2008non}. This scenario arises theoretically under certain conditions in models such as the Kitaev fermion chain \cite{kitaev2001unpaired} in which the encoded information is protected against fluctuations in local electric potentials and can be relatively easy to isolate against external sources of charge. The Kitaev chain provided inspiration for looking at non-local encoding of qubits in the mathematically equivalent spin chain with nearest-neighbour interactions known as the anisotropic XY spin chain model, where a related topologically non-trivial phases can appear \cite{schultz1964ising,greiter20141d,you2014encoding,levitov2001quantum}. Using spins (or pseudo-spins) to realise such states is attractive since there are several physical systems which can simulate spins, e.g., superconducting qubits, trapped ions, quantum dots, and impurities in silicon. Unfortunately spin chains are coupled to a gapless bosonic environment and the ground state qubit is not protected even against local uncorrelated noise. Consequently such noise can couple the two ground states and hence scramble the encoded information. 

In this paper we explore a spin chain combining local and non-local spin-spin interactions and find that it admits a phase with a fully protected two-fold degenerate ground state manifold and gapped from the higher excited states. We show that, in sharp contrast to the XY chain, the quantum information encoded in these states will remain uncoupled from local external noise. We attribute this protection to the strong suppression of all matrix elements between the ground states for all possible local spin couplings and all the two-body couplings that arise from fluctuations in the Hamiltonian parameters. We demonstrate that the protection persists in a range of parameters remarkably even in short chains of 8-14 qubits which are subject to strong finite size effects. We explore the boundaries of the phase and discover that it is separated from the XY model by a phase transition, controlled by the anisotropy parameter. Finally, employing a ground-state averaged entanglement entropy as a measure, we find that it acquires a universal value within the new phase, distinct from the universal value of the XY model. 

The phase we find is considered here, while other schemes for encoding with different complex spin systems have been of theoretical interest recently\cite{LossPhysRevB.86.205412,kapitPRA2015passive, milmanPRL2007topo}. Owing to its protection, it avoids many of the shortcomings of the XY model in encoding quantum information in the ground state manifold. While a full discussion of its realisation is beyond the scope of this investigation, it is inspired by the types of Hamiltonians which typically arise for superconducting qubit-cavity systems. Superconducting qubits may prove an attractive venue due to the precision with which the inter-qubit coupling can be controlled
\cite{van2007controllable,hime2006solid,niskanen2007quantum,mckay2015tunable_exp,chen2014tunable_exp} and due to progress in techniques for preparing, simulating and measuring correlated qubit states \cite{dicarlo2010preparation,shai2015cooling,wallraff2015simulating,heras2014digitalsimulation,barends2015adiabatic}. 
Nevertheless, the model is general and can be potentially realised in other physical systems.  The protection of the ground state manifold to all local spin perturbations suggests a certain robustness of the phase against additional couplings that may arise in an actual physical system. Such perturbations are suppressed by the gap to the first excited state as predicted by second order perturbation theory. 

\section{A model of hybrid local and non-local interaction} We consider the following coupled spin-chain model 
\begin{align}
H_{S} & =-\frac{1}{2}\sum_{j=0}^{N-2}\left[\left(t+\Delta\right)\sigma_{j}^{x}\sigma_{j+1}^{x}+\left(t-\Delta\right)\sigma_{j}^{y}\sigma_{j+1}^{y}\right]\nonumber \\
 & +\frac{\lambda_{FF}}{2}\sum_{i,j=0}^{N-1}\left[\sigma_{i}^{-}\sigma_{j}^{+}+\sigma_{i}^{+}\sigma_{j}^{-}\right]-\frac{\mu}{2}\sum_{j=0}^{N-1}\sigma_{j}^{z}.\label{eq:spin_ham}
\end{align}
Here nearest-neighbours (NN) interaction $t>0$ can be
realised for example in a system of superconducting qubits coupled
to each other capacitively or inductively \cite{you2014encoding,levitov2001quantum}.
A route to realising a general anisotropic coupling was discussed
recently \cite{kapit2015universal} (anisotropic interactions are not critical for realising our proposed phase, but play an essential role in simulating the XY model). 
Related spin-spin interactions
have been successfully realised also in trapped ions \cite{chang2010frustrated}.
Non-local ``flip-flop'' interactions $\lambda_{FF}>0$ arise naturally in the context of superconducting circuit QED setups, when the qubits
are all strongly coupled to a common superconducting resonator and
are well detuned from its resonance frequency \cite{Blais2004cavityqed,zheng2000efficient}. For a homogeneous case the interactions can be expressed with a total pseudo-spin
operator $S^{(x,y,z)}=\sum_{i=1,N}\sigma_{i}^{(x,y,z)}$ as $\sum_{i,j=1,N}\left[\sigma_{i}^{-}\sigma_{j}^{+}+\sigma_{i}^{+}\sigma_{j}^{-}\right]=S^{-}S^{+}+S^{-}S^{+}=2({\bf S^{2}}-S_{z}^{2})$.
The magnetic field $\mu$ is usually taken to be the detuning between the qubit transition
frequency and the drive or measurement tones. Spin chains with negative NN and positive next-to-NN interactions were considered
recently in quantum magnetism \cite{furusaki2015nnn}.

We perform an exact diagonalisation study of the Hamiltonian $H_S$ and identify a new phase of the spin chain which is most pronounced in the isotropic case of the NN couplings $\Delta\ll t,\lambda_{FF}$ and close to $\mu=0$, \emph{i.e.}, in the rotating frame. In the following we characterize the main properties of this phase and of its phase boundaries.

\section{Characterization of the new phase} 
\begin{figure}[h]
\begin{centering}
\includegraphics[scale=0.3]{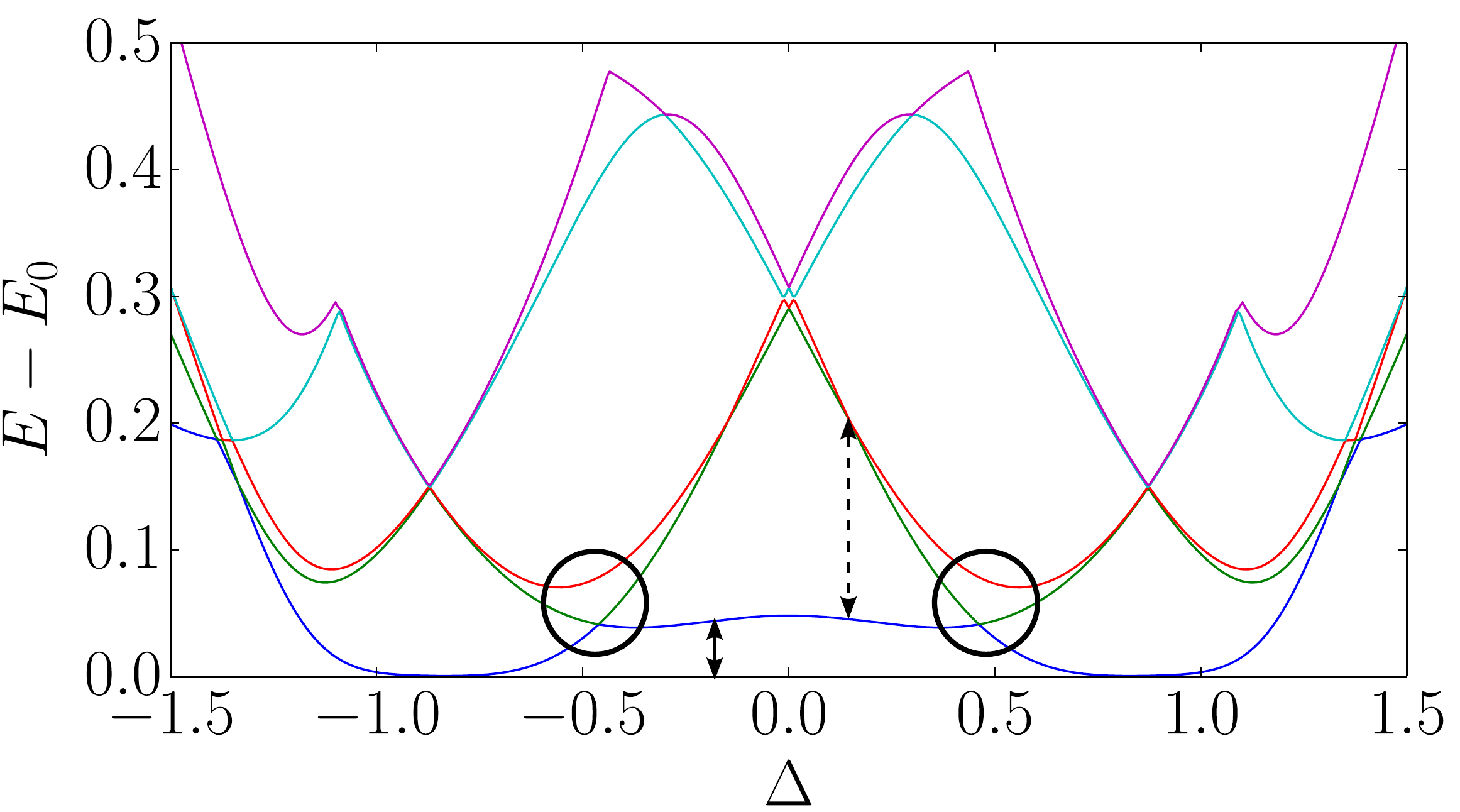}
\par\end{centering}

\begin{centering}
\includegraphics[scale=0.3]{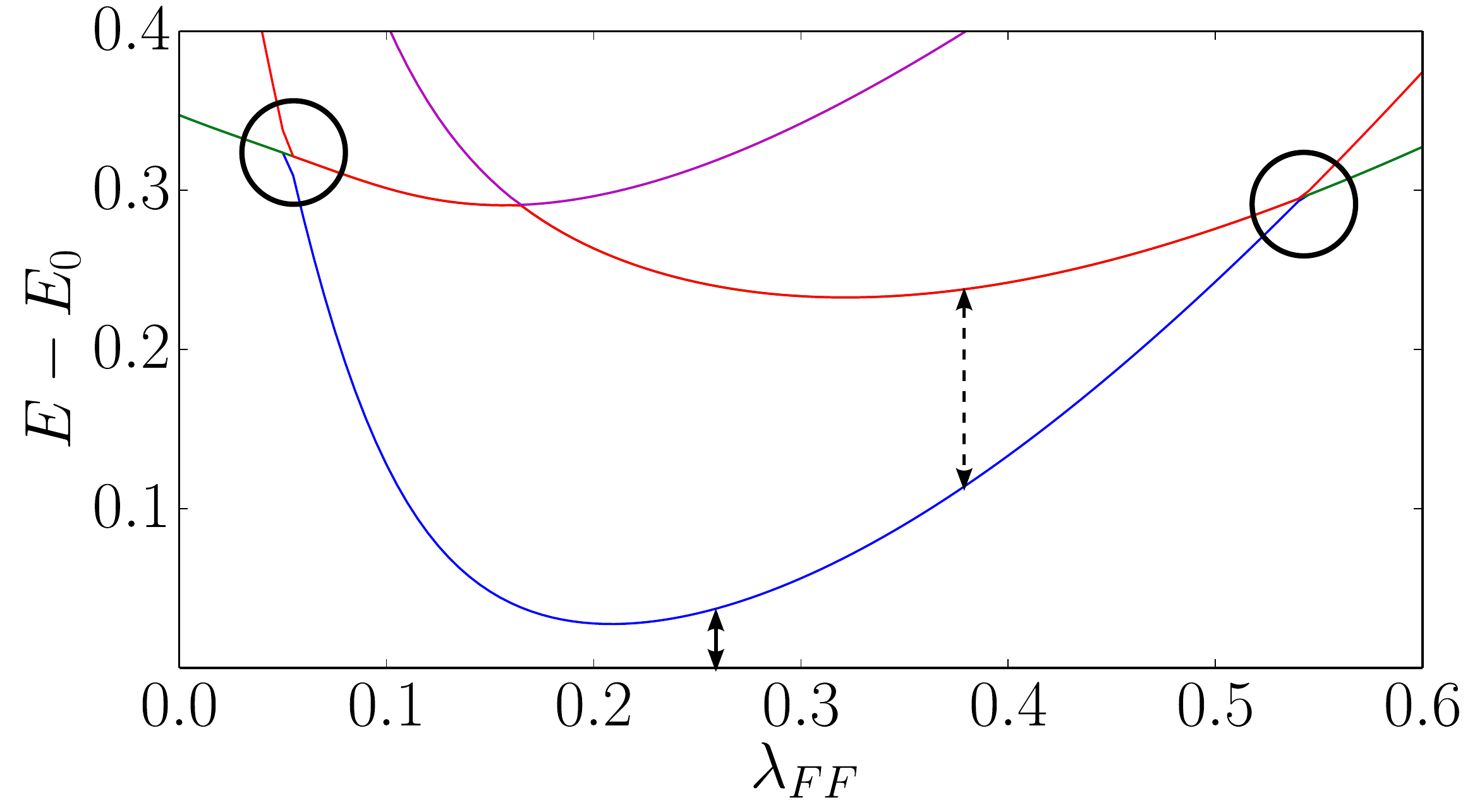}
\par\end{centering}

\caption{\label{fig:specdiff} Low lying energy eigenvalues of the spin chain as a function of parameters show transition into the protected doublet phase shown as a function of decreasing anisotropy $\Delta$ (upper) or increasing non-local interactions (bottom) as the bulk gap closes and reopens at the transition point (black circle) leaving a quasi-degenerate ground state for a finite range. The plots are of differences between the $n$th lowest energy level, $E$, (for $n=1,2,3,4,5$) and the lowest energy level $E_{0}$ for an $N=8$ spin-chain with $\mu=0$, $t=1$ as a function of $\Delta$ (top, $\lambda_{FF}=0.15$) and of $\lambda_{FF}$ (bottom, $\Delta=0$). The solid arrows shows the splitting between nearly degenerate ground-states and the dotted arrows shows the bulk gap. Plots are differentiated by colors/shades for clarity and $t=1$ for all cases. The cases of $N=10,12$ are shown in appendix~\ref{app:longer-chains}.}
\end{figure}

\begin{figure}[h]
\begin{centering}
\includegraphics[scale=0.3]{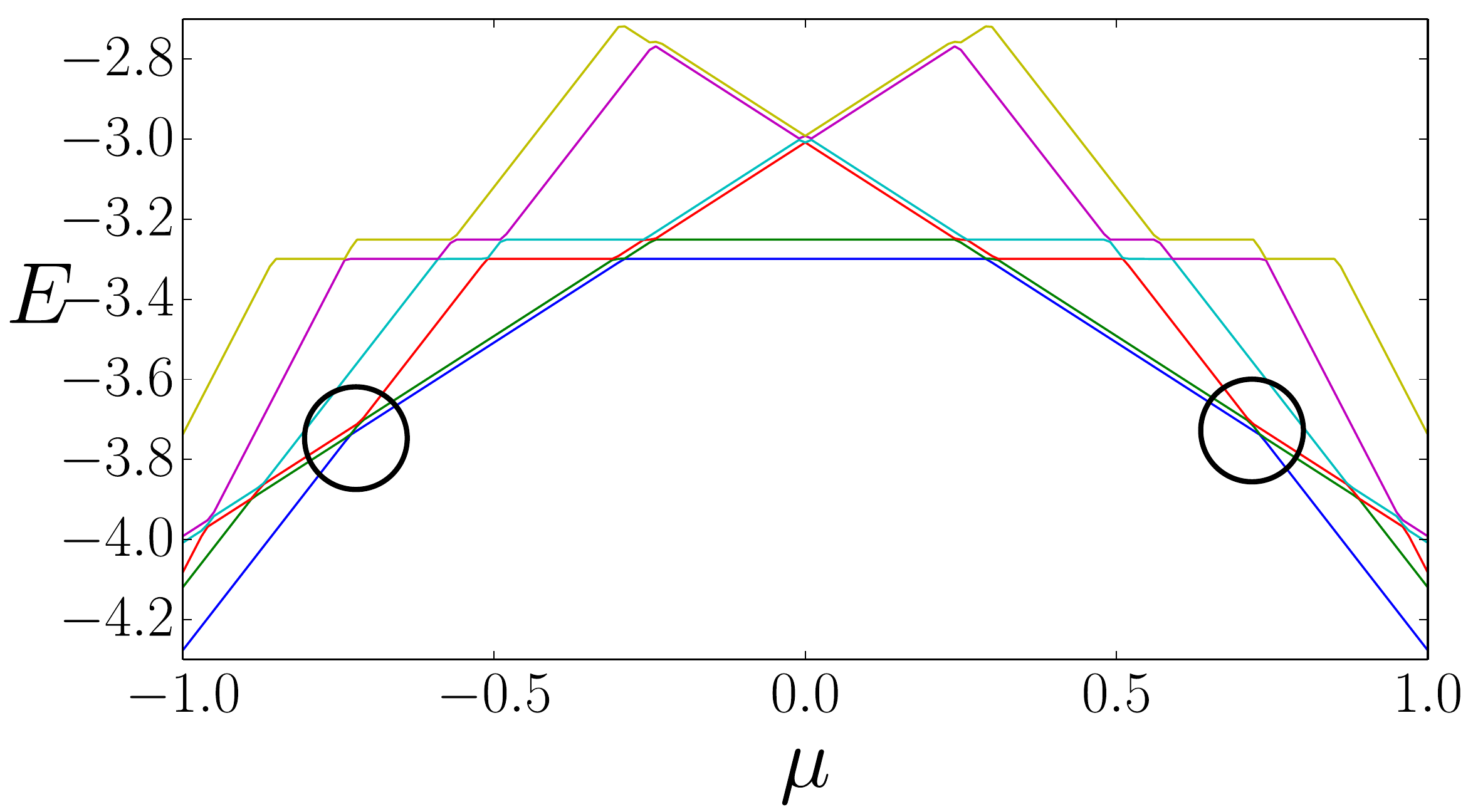}
\par\end{centering}

\caption{\label{fig:specabs} Low lying energy eigenvalues of the spin chain as a function of the spin polarising energy $\mu$ are shown to have different behaviour inside and outside the protected phase. The $n$th
lowest energy level, $E$, (for $n=0,1,2,3,4,5$) for an $N=8$ spin-chain with $\Delta=0$, $\lambda_{FF}=0.15$ as a function of $\mu$.  A transition from a single ground state into a gapped
doublet appears at a certain critical value (black circle) even though the pairing energy $\Delta$ is zero and there are no Kitaev chain zero modes. Plots are differentiated by colors/shades for clarity and $t=1$ for all cases.}
\end{figure}

The first property of interest is the appearance of a quasi-degenerate doublet of ground
states which is separated by a gap from the excited states. This occurs for a finite chain with open boundary conditions  when the non-local coupling strength $\lambda_{FF}$, which couples each spin to $N-1$ other spins, is of the order of $t/N$ and within a finite range of the anisotropy $\Delta$ close to $\Delta=0$ (see Fig.~\ref{fig:specdiff}). For larger values of $\Delta$ it can be seen that the bulk gap closes and reopens at the
boundaries of the protected region (see Figs. \ref{fig:specdiff},\ref{fig:specabs}),
suggesting that the non-local interactions have introduced a distinct, correlated phase. As is apparent from the figure, the XY model also possesses a doublet of ground states. However, as we now turn to discuss, it is distinguished by its degree of protection to external perturbations.

\begin{figure}
	\begin{centering}
		\includegraphics[scale=0.2]{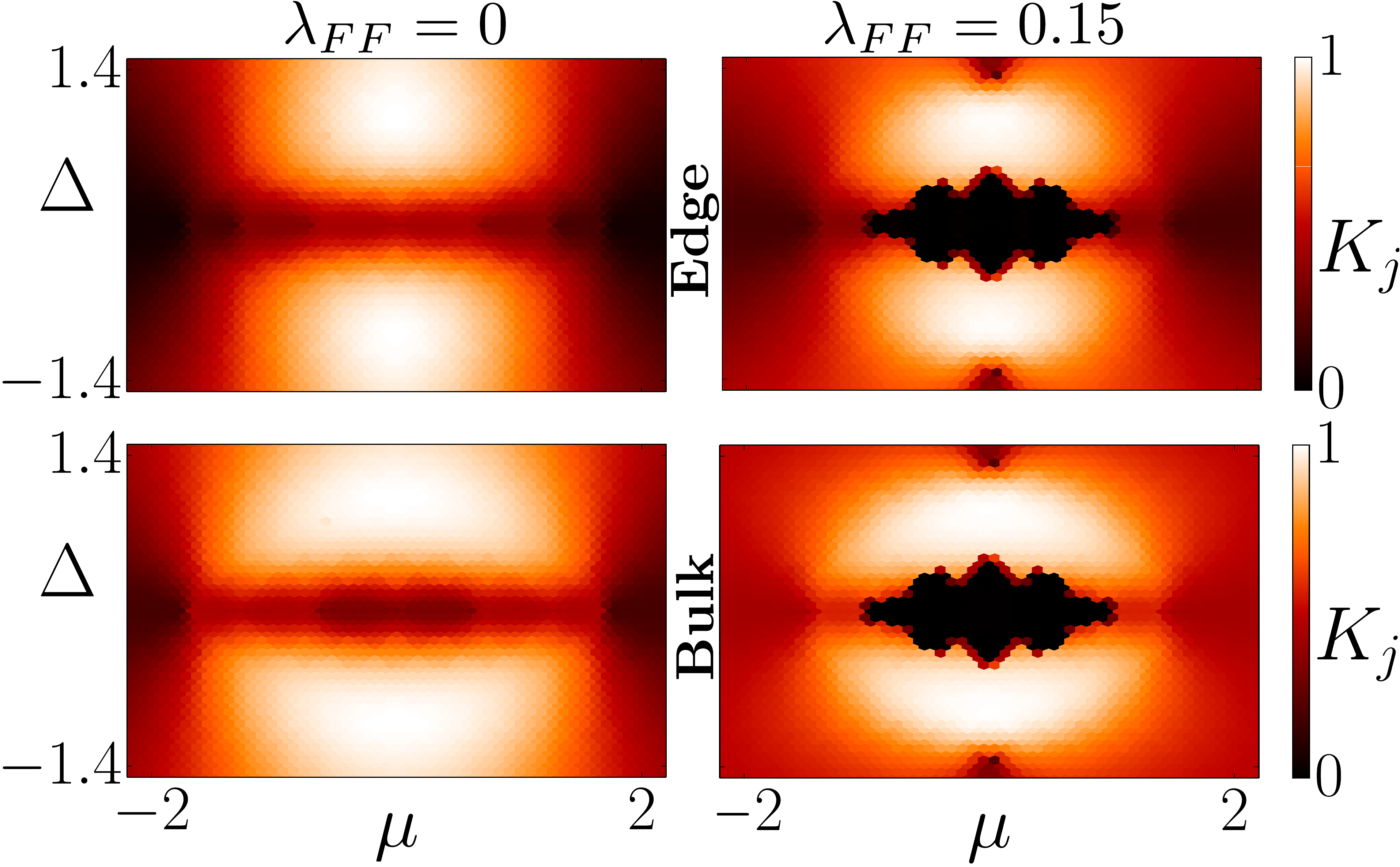}
		\par\end{centering}
	
	\caption{\label{fig:total-coupling}The effect of introducing cavity-mediated
		(non-local) interactions on the coupling of the ground states. The
		total coupling $K_{j}=\left|\left\langle 0\right|\sigma_{j}^{x}\left|1\right\rangle \right|^{2}+\left|\left\langle 0\right|\sigma_{j}^{y}\left|1\right\rangle \right|^{2}+\left|\left\langle 0\right|\sigma_{j}^{z}\left|1\right\rangle \right|^{2}$at
		the end of the chain (upper) vs. at a bulk (bottom) site without (left)
		vs. with (right) additional non-local interactions $U\sigma_{i,+}\sigma_{j,-}$for
		all $i,j$ and $\lambda_{FF}=0.15$ on an $N=8$ chain (similar features
		appear for $N=6$ and $N=10$). When the non-local interactions are
		added a region of only very weak total coupling ($\sim10^{-3}-10^{-4}$)
		opens at the central region of the parameters space. We set $t=1$ for all cases.}
\end{figure}

The difference between the (non-interacting) XY phase and the new phase manifests most strikingly in the matrix elements of the local spin operators between the degenerate ground states and their dependence on the system parameters. These coupling patterns for the XY chain are presented in Fig.~\ref{fig:total-coupling} (left), and are more or less constant and seem fairly unremarkable. The efficient coupling between the two ground states, seen in Fig.~\ref{fig:total-coupling} and Fig.~\ref{fig:gsc}, can be understood in the case of $\lambda=0, \Delta=t$ within the quantum Ising model with a transverse field $\mu$. The two ground states are symmetric and anti-symmetric superpositions of $|\rightarrow, \rightarrow,..\rangle$ and $|\leftarrow,\leftarrow,..\rangle$ where $|\rightarrow(\leftarrow)\rangle$ denote eigenstates of $\sigma^x$. Hence it is easy to see that a local operator $\sigma^x_i$ at site $i$ can change one ground state into the other. In contrast, when turning $\lambda_{FF}$ on all three spin coupling strengths are suppressed within the new correlated phase, so the ground state manifold is completely protected against external perturbations of the type $\sigma^{x,y,z}$. This manifests as the striking black regions in Fig.~\ref{fig:total-coupling} (r.h.s.) which are calculated for open boundary conditions\footnote{Central site pertrurbations do couple the ground states however this coupling diminishes with the chain length and vanishes altogether for the case of periodic boundary conditions.}. Due to this complete cancellation in all spin directions, also other choices of basis states must remain decoupled. Interestingly the protected phase survives in some finite region of parameters even in the presence of a moderate amount of $10\%$ disorder in either site or NN coupling energies (see appendix~\ref{app:dis}). It is also worth noting that the two-body perturbations, most likely to be introduced by fluctuations in the model parameters\cite{you2014encoding} such as $t,\lambda$ here do not couple the two degenerate ground states. This is in contrast to what is seen in related models such as the Majumdar-Ghosh model\cite{majumdar1969next} where interactions extend only to next-nearest-neighbours.

We note that the splitting of near-degenerate ground states is much smaller than the other energy
scales in the system. By inspecting chains of different lengths $N=6,8,10,12,14,16$ we see that
the residual doublet splitting decreases with system size and most features sharpen. This suggests that the residual splitting is due to
the finite size of the chain and it trends towards an exact degeneracy for the longer chains (see appendix~\ref{app:longer-chains}). This phase indeed requires a long-range interaction in order to appear which however can fall off gradually (see appendix~\ref{app:soft}).
The transition into this phase can be explored both from the direction of reducing the anisotropy
$\Delta$ or from the direction of reducing the spin polarizing energy $\mu$. In the former ($\Delta$) first the
degeneracy splits and the upper state (Fig. \ref{fig:specdiff}, blue
curve) switches place with another excited state which descends from
above (Fig. \ref{fig:specdiff}, green curve). In the latter ($\mu$)
a transition from a single ground state into a gapped doublet appears
at a certain critical value (see Fig. \ref{fig:specabs}), while $\Delta$ is zero, distinguishing it from the XY
phase. Increasing $\mu$, positively or negatively, eventually
leads to the closing of the gap and transition into a more polarised 
and less correlated states with a single ground state (see circles). 

\begin{figure}
\begin{centering}
\includegraphics[scale=0.3]{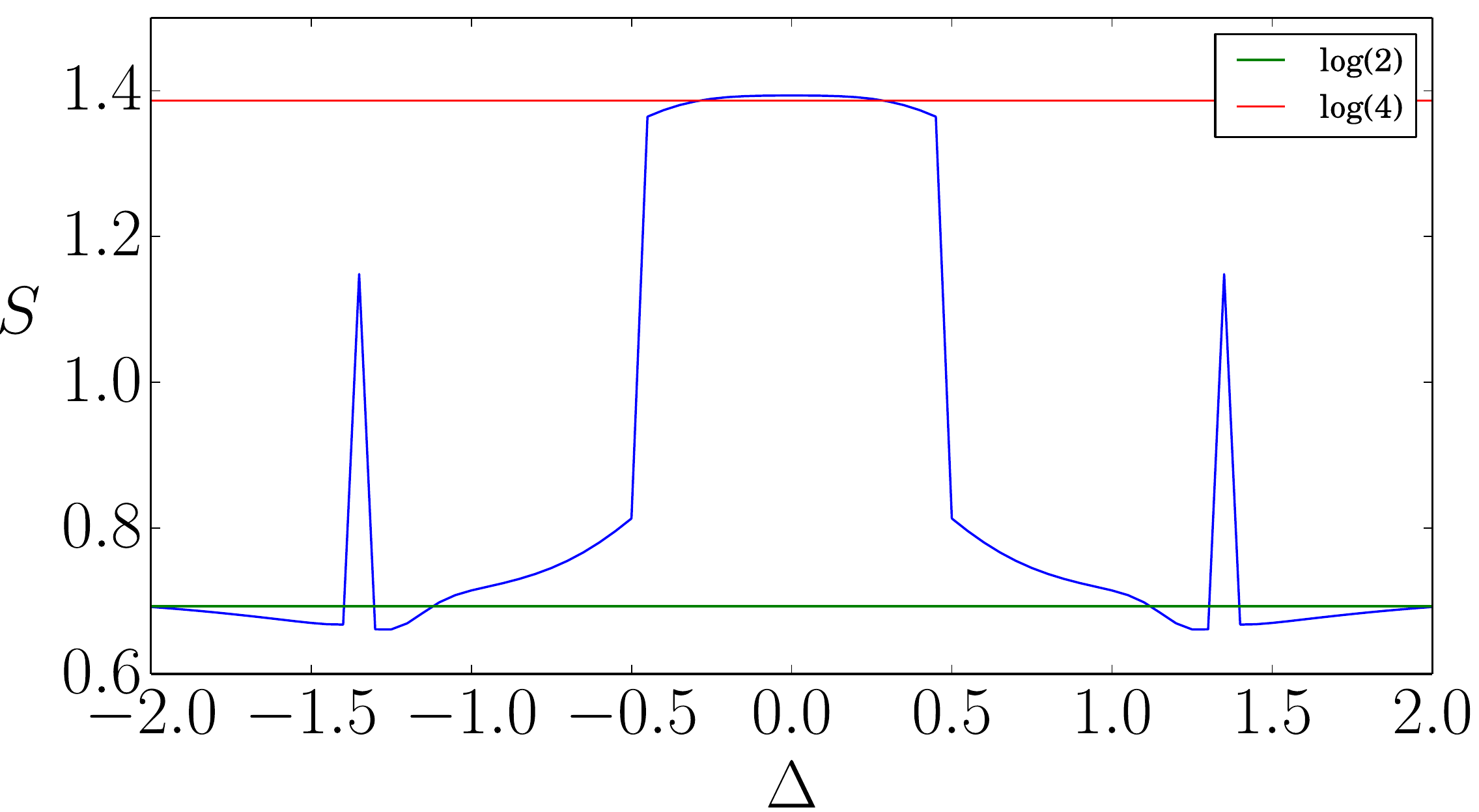}
\par\end{centering}

\caption{\label{fig:ent}The correlations within the spin chain increase sharply as the system enters the protected phase. This is demonstrated with the entanglement entropy (EE) for partitioning the
system into two equal chains, presented as function of the parameter
$\Delta/t$. Using the same parameters as in Fig. \ref{fig:specdiff},
the EE presents a $2\ln(2)$ plateau within the new phase, which rapidly
drops at the phase transition. We set $t=1$ for all cases.}
\end{figure}

Finally, we explore the bipartite entropy (BE)
of the ground state which is obtained by splitting the chain to two equal left and right parts and is defined as the von Neumann entropy of the reduced density matrix of one side. In the Kitaev phase, for the mixed state of the
two ground states $\rho=\frac{1}{2}(\left|0\right\rangle \left\langle 0\right|+\left|1\right\rangle \left\langle 1\right|)$
it is quantised at the value of $\ln(2)$ due to the topological
nature of the states and their distinct parity symmetries. Here it
is modified by the interactions, see Fig. \ref{fig:ent}, but still
approaches the value of $\ln(2)$ asymptotically as $\Delta$ is
increased. Interestingly, as $\Delta$ is decreased and the system
enters the protected phase the same measure of EE jumps to a higher
value of $\ln(4)$. It is plausible that the enhanced degree of correlations within the ground states, generated by long-range interactions, make it less sensitive to local perturbations.

Due to the small splitting between
the ground states, the system will be found in a thermal mixture
state. Initial pure state preparation is conceivable by preferentially
driving the system to an excited state from which it decays the ground
states (optical pumping) or by adiabatically turning on the interactions.
We observe that the expectation values of $\langle0|\sigma_{j}^{z}|0\rangle$
and $\langle1|\sigma_{j}^{z}|1\rangle$ are generally different
in most of parameter space. Therefore preparation of a ground
state followed by a local measurement of $\sigma_{i}^{z}$ can reveal
the relaxation time. By crossing the protected phase boundary in the
$\Delta$ direction, a significant difference should be observed as
the states become coupled by external perturbations.

\section{Comparison to the Kitaev model} It is instructive to compare the situation to the fermionic Kitaev chain model where the qubit is encoded in the state of two Majorana zero-modes (MZMs) \cite{majorana1937symmetric,sarma2015majorana,read2000paired} and realisation of MZMs in fermionic systems has been discussed extensively \cite{fu2008superconducting,sau2010generic,oreg2010helical,ginossar2014microwave,ginossar2015fermion}, with signatures compatible with MZMs observed recently \cite{mourik2012signatures,nadj2014observation}. Although robust against certain types of local noise sources, some processes are predicted to still cause decoherence of the encoded qubit. The system is mostly sensitive to a fermion bath coupling at low energies. Such baths could couple mostly to the edge of the chain \cite{ho2014decoherence,goldstein2011decay,schmidt2012decoherence,rainis2012majorana,ng2014decoherence}. As noted above the XY model was considered for potential realisation for the Kitaev chain as it can be straightforwardly simulated in highly controllable quantum systems such as superconducting qubits \cite{you2014encoding}. While formally equivalent to the Kitaev chain, the two models were understood to be physically different due to the non-locality of the mathematical mapping between them (see appendix~\ref{app:bdg}). Consequently, typical external perturbations affecting the XY spin chain model would act essentially different as compared to the perturbations affecting the fermionic Kitaev chain. In the former perturbations originate from the coupling to a bosonic reservoir while in the latter from the coupling to a fermion reservoir. As a consequence bosonic bath coupled locally to the spins at any point of the XY chain could directly couple the two encoded qubit ground states. Hence the XY model with only local interactions between the spins ($\lambda_{FF}=0$) shows virtually no protection of the ground state manifold but a large scale realisation of the XY model would be useful for the study of MZM decoherence.

\begin{figure}
	\begin{centering}
		\includegraphics[scale=0.2]{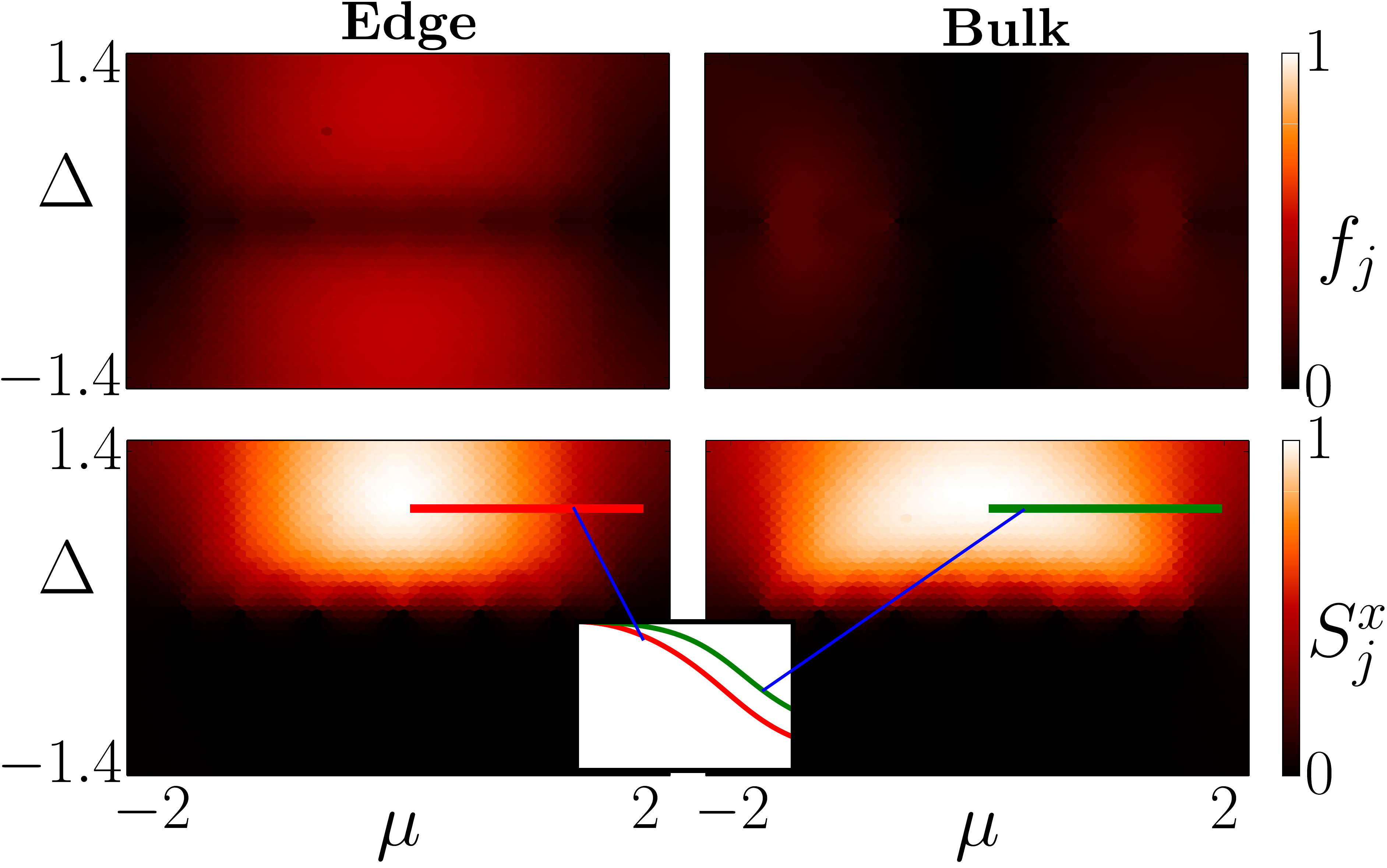}
		\par\end{centering}
	
	\centering{}\caption{\label{fig:gsc} Degree of ground-state coupling induced by fermion
		injection perturbation (upper) vs. coupling induced by a spin-flipping
		perturbation in the analogue spin model (bottom). Transition strengths
		$f_{j}$ for Kitaev model (top) and $S_{j}^{x}$ for spin model between
		ground-states for edge ($j=0$, left) and bulk ($j=1$, right) site
		of an 8-site chain. For the Kitaev model, it can be seen that the
		transition strength is high ($\sim0.5$) in much of the parameter
		range for the edge site and low ($\sim0$) in much of the parameter
		range for the bulk site. For the spin model, it can be seen for both
		that the transition strength is high for much of the positive range
		of $\Delta$, but close to 0 for much of the negative range. The difference
		between edge and bulk is much less pronounced for the spin model;
		for the bulk site, the high-strength region is slightly larger. This
		is shown also in at the positions of the green and red lines, chosen
		at $\Delta=1$ (see inset). We set $t=1$ for all cases.}
\end{figure}

When the femionic chain couples to a fermionic bath, a gauge invariant measure for the induced transitions between degenerate ground states is given by the following quantity
\begin{equation}
f_{j}=|\langle0| a_{j}^{\dagger}| 1\rangle|^{2}+|\langle0| a_{j}| 1\rangle|^{2},\label{eq:A_j}
\end{equation}
at each site $j$, where $|0\rangle$, $|1\rangle$ are the
lowest two states of the system and $a_{j}$, $a_{j}^{\dagger}$ are fermionic annihilation and
creation operators of the spinless chain at site $j$, respectively. The quantity $f_{j}$, as defined in Eq.~\ref{eq:A_j}, has been calculated
numerically for an edge site ($j=0$) and a bulk site ($j=1$) for
an 8-site Kitaev chain. $f_{0}$ and $f_{1}$ are shown in Fig. \ref{fig:gsc}.
This figure also shows a clear edge effect in the context of the
Kitaev chain: it appears much easier to couple the two ground-states
over most of the parameter range at the edge site than in the middle
of the chain. This is consistent with the Majorana fermion picture:
the MFs constituting the MZM are individually localised on the edge
sites, rendering them easier to affect there. In fact, it can be seen that the edge site transition is strongest
at the ideal points ($\mu=0$, $\Delta=\pm1$), where localisation
is perfect; away from this point, the MFs decay into the bulk of the
chain and $f_{0}$ is reduced. Conversely, the bulk site transition
$f_{1}$ is 0 at the ideal point, where it can have no effect on the
MFs, and gradually increases away from this point as the MFs begin
to decay into the bulk. This is consistent
with the conductivity measured in an experiment like the one in \cite{nadj2014observation}.
Importantly, this response could be observed in the spin-chain
by recognising that $\left(a_{j}^{\dagger},a_{j}\right)$ translate
into the fictitious spin perturbations $\left[\prod_{i=0}^{j-1}-\sigma_{j}^{z}\right]\left(\sigma_{j}^{+},\sigma_{j}^{-}\right)$.
These are experimentally accessible for $j=0$ by perturbing
with $\left(\sigma_{0}^{+},\sigma_{0}^{-}\right)$, respectively,
and $j=1$, by perturbing with $-\sigma_{0}^{z}\left(\sigma_{1}^{+},\sigma_{1}^{-}\right)$.
Extracting the response of the spin-system is a test
for the Kitaev chain behaviour in the spin system. 

As noted above, local spin perturbations can strongly couple these ground states in the XY model. This can be seen and studied through the quantity $S_{j}^{x}$, defined in Eq. \ref{eq:S_j},
\begin{equation}
S_{j}^{x}=|\langle0|\sigma_{j}^{x}|1\rangle|^{2}\label{eq:S_j}
\end{equation}
which has been calculated numerically for parameters in the range $-2\leq\mu<2$
and $-1\leq\Delta<1$ for an edge site ($j=0$) and a bulk site ($j=1$)
for an 8-site spin chain. $S_{0}^{x}$ and $S_{1}^{x}$ are shown
in Fig. \ref{fig:gsc} (bottom). While for negative $\Delta$ the states appear uncoupled for $\sigma_{j}^{x}$, they are coupled for the $\sigma_{j}^{y}$ in this region, namely the graph is symmetrically inverted. For $\sigma_{j}^z$ the states are uncoupled however overall the states are unprotected for the spin XY model. 

Fig. \ref{fig:gsc} also shows some edge effect for the local spin operator $\sigma_{j}^{x}$. The difference in the response between the edge and bulk for spins
can be shown to be directly related to the edge-localised Majorana wave function in the analogue fermionic chain, using the Bogoliubov de-Gennes formalism (see appendix~\cref{app:bdg,app:edge-sig,app:asym-response}). 

In conclusion, the results indicate existence of a phase with highly correlated spin ground state doublet in the presence of non-local interactions. We studied the properties of this phase and discovered a remarkable resilience to perturbations that extends beyond the protection offered by the XY model. These properties are suggestive of a topological phase which is distinct from the phase of the XY model. This phase may offer resource-efficient quantum state encoding with enhanced protection against uncorrelated local perturbations. The non-local interactions appear naturally in circuit quantum electrodynamics but may be realized in other systems with effective long range exchange interactions.

\begin{acknowledgements} 
EGr acknowledges  support  from  the  Israel  Science  Foundation  under  grant  No. 1626/16,   the  European  Union?s  Seventh  Framework  Programme (FP7/2007-2013) under grant No.  303742, and the United States-Israel Binational Science Foundation  under  grant  No. 2014345.EGi acknowledges
support from EPSRC (EP/L026082/1). EGr and EGi acknowledge support
from the Royal Society International Exchanges programme, Grant No.
IE121282.
\end{acknowledgements}

\appendix

\section{Bogoliubov de-gennes Method}\label{app:bdg}

\subsection*{Quasiparticle operators in the local operator basis}

\begin{figure}[h]
	\begin{centering}
		\includegraphics[scale=0.35]{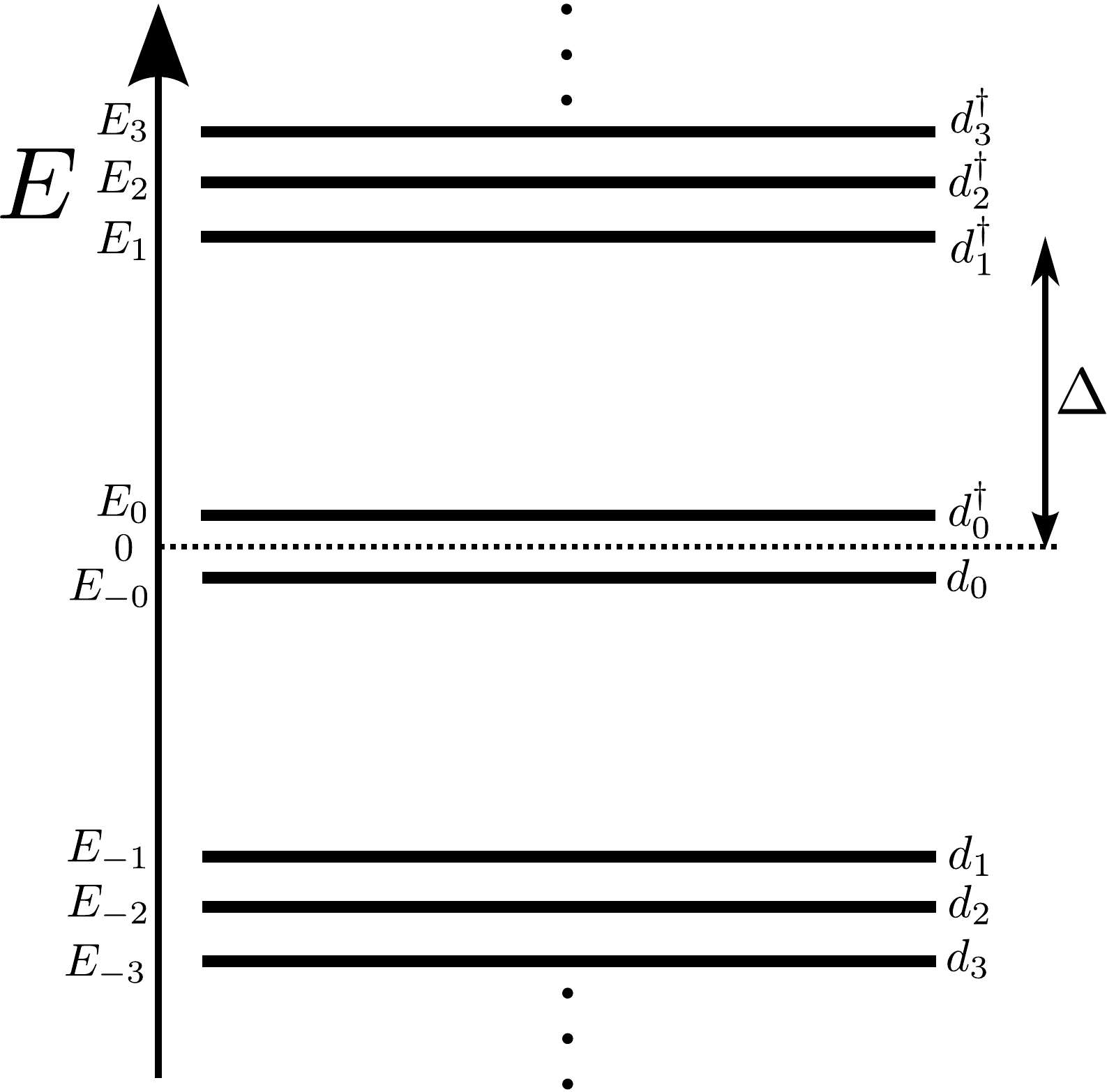}\caption{\label{fig:schematic}A schematic of the quasi-particle spectrum in	the Kitaev chain model. The quasiparticles are created ($d^\dagger_n$) with a minimal energy of $E_1=\Delta$.}
		\par\end{centering}
\end{figure}

For completeness we review the BdG method which is used to obtain Fig. \ref{fig:gsc}. The Kitaev Hamiltonian, defined with fermion creation and annihilation operators $a_j^\dagger,a_j$ for site $j$ is
\begin{align}
& H=-\mu\sum_{j=0}^{N-1}\left(a_{j}^{\dagger}a_{j}-\frac{1}{2}\right)\nonumber \\
& +\sum_{j=0}^{N-2}\Delta a_{j}a_{j+1}+\Delta^{*}a_{j+1}^{\dagger}a_{j}^{\dagger}\label{eq:kit_ham}\\
& -\sum_{j=0}^{N-2}t\left(a_{j}^{\dagger}a_{j+1}+a_{j+1}^{\dagger}a_{j}\right)\nonumber 
\end{align}
can be written as
\[
H=\mathcal{C^{\dagger}\mathcal{H}}\mathcal{C}
\]
where
\[
\mathcal{C}=\left(\begin{array}{c}
a_{0}\\
\vdots\\
a_{N-1}\\
a_{0}^{\dagger}\\
\vdots\\
a_{N-1}^{\dagger}
\end{array}\right)
\]
and $\mathcal{H}$ is a $2N$-by-$2N$ (where $N$ is the number of
chain sites) hermitian matrix which depends on $\mu$, $t$ and $\Delta$.
Let the $2N$ eigenvalues of $\mathcal{H}$ be written $E_{\pm n}$
for $0\leq n<N-1$ and be indexed such that $E_{a}\leq E_{b}$ if
$a<b$. These eigenvalues represent the single-particle spectrum of
the Kitaev model, symmetric about $E=0$, where $E_{-n}=-E_{n}$.
The corresponding eigenvectors, $d_{\pm n}^{\dagger}$ where $d_{-n}^{\dagger}=d_{n}$,
are the creation and annihilation operators for the elementary excitation
quasiparticles, expressed in the $a_{n}$, $a_{n}^{\dagger}$ basis.
This picture is illustrated in fig. \ref{fig:schematic}.

The unitary transformation, $U$, which diagonalises
$\mathcal{H}$ with $U^{\dagger}\mathcal{H}U$ can be written as
\[
U=\left(\begin{array}{cccccc}
u_{0}^{(-N+1)} & \dots & u_{0}^{(-0)} & u_{0}^{(+0)} & \dots & u_{0}^{(N-1)}\\
\vdots & \vdots & \vdots & \vdots & \vdots & \vdots\\
u_{N-1}^{(-N+1)} & \dots & u_{N-1}^{(-0)} & u_{N-1}^{(+0)} & \dots & u_{N-1}^{(N-1)}\\
v_{0}^{(-N+1)} & \dots & v_{0}^{(-0)} & v_{0}^{(+0)} & \dots & v_{0}^{(N-1)}\\
\vdots & \vdots & \vdots & \vdots & \vdots & \vdots\\
v_{N-1}^{(-N+1)} & \dots & v_{N-1}^{(-0)} & v_{N-1}^{(+0)} & \dots & v_{N-1}^{(N-1)}
\end{array}\right)
\]
and so the fermionic quasiparticle operators can be found by the matrix-vector
multiplication
\[
\left(\begin{array}{c}
a_{0}\\
\vdots\\
a_{N-1}\\
a_{0}^{\dagger}\\
\vdots\\
a_{N-1}^{\dagger}
\end{array}\right)=U\left(\begin{array}{c}
d_{N-1}\\
\vdots\\
d_{0}\\
d_{0}^{\dagger}\\
\vdots\\
d_{N-1}^{\dagger}
\end{array}\right).
\]
It is worth noting that, by comparing $d_{\pm n}^{\dagger}=\sum_{i=0}^{N-1}u_{i}^{(\pm n)}a_{n}+v_{i}^{(\pm n)}a_{n}^{\dagger}$
with $d_{\pm n}=d_{\mp n}^{\dagger}=\sum_{i=0}^{N-1}u_{i}^{(\pm n)*}a_{n}^{\dagger}+v_{i}^{(\pm n)*}a_{n}$,
it can be seen that $u_{i}^{(\pm n)},v_{i}^{(\pm n)}=v_{i}^{(\mp n)*},u_{i}^{(\mp n)*}$.

\subsection*{Ground-state coupling for fermions}

To contrast the simplicity of applying BdG to the fermion system with
the complexity of doing the same to the spin system, we first turn
our attention to the quantity $f_{j}=\mid\langle0\mid a_{j}^{\dagger}\mid1\rangle\mid^{2}+\mid\langle0\mid a_{j}\mid1\rangle\mid^{2}$,
which we write as 
\[
f_{j}=\mid f_{j}^{+}\mid^{2}+\mid f_{j}^{-}\mid^{2}
\]
where $\left(f_{j}^{+},f_{j}^{-}\right)=\langle0\mid\left(a_{j}^{\dagger},a_{j}\right)\mid1\rangle$.

The operators of interest are
\[
(a^{\dagger},a)_{i}=\sum_{n=0}^{N-1}\left(u,v\right)_{i}^{(-n)}d_{n}+\left(u,v\right)_{i}^{(+n)}d_{n}^{\dagger}
\]
and so $\left(f_{j}^{+},f_{j}^{-}\right)=\langle0\mid\sum_{n=0}^{N-1}\left(u_{j}^{(-n)},v_{j}^{(-n)}\right)d_{n}d_{0}^{\dagger}+\left(u_{j}^{(+n)},v_{j}^{(+n)}\right)d_{n}^{\dagger}d_{0}^{\dagger}\mid0\rangle$.
Since the $d_{\pm n}^{\dagger}$ (for all $0\le n<N-1$) are eigenoperators
of $\mathcal{H}$ and $\mid0\rangle$, $\mid1\rangle$ are eigenstates
of $\mathcal{H}$, the only non-vanishing term in the sum is the one
containing $d_{0}d_{0}^{\dagger}$. Thus,
\[
f_{j}^{+},f_{j}^{-}=u_{j}^{(-0)},v_{j}^{(-0)}
\]

An alternative way to achieve the same expression, which involves
the BCS correlators, is to instead use the substitution $d_{0}^{\dagger}=\sum_{n=0}^{N-1}u_{n}^{(+0)}a_{n}+v_{n}^{(+0)}a_{n}^{\dagger}$.
Now,
\begin{eqnarray*}
&& \left(f_{i}^{+},f_{i}^{-}\right) =  \\
&&\langle0\mid\sum_{j=0}^{N-1}u_{n}^{(+0)}\langle0\mid(a_{i},a_{i}^{\dagger})a_{j}\mid0\rangle+v_{n}^{(+0)}\langle0\mid(a_{i,}a_{i}^{\dagger})a_{j}^{\dagger}\mid0\rangle\\
&& =\sum_{j=0}^{N-1}u_{j}^{(+0)}\left(F_{i,j}^{\prime},C_{i,j}\right)+v_{j}^{(+0)}(C_{i,j}^{\prime},F_{i,j})
\end{eqnarray*}
where
\begin{eqnarray*}
	\left(C_{i,j},F_{i,j}\right) & = & \langle0\mid a_{i}^{\dagger}\left(a_{j},a_{j}^{\dagger}\right)\mid0\rangle\\
	& = & \sum_{n=0}^{N-1}v_{i}^{(-n)}(u_{j}^{(+n)},v_{j}^{(+n)})
\end{eqnarray*}
and
\begin{eqnarray*}
	F_{i,j}^{\prime},C_{i,j}^{\prime} & = & (F_{j,i}^{*},\delta_{ij}-C_{j,i})\\
	& = & \langle0\mid a_{i}\left(a_{j},a_{j}^{\dagger}\right)\mid0\rangle\\
	& = & \sum_{n=0}^{N-1}u_{i}^{(-n)}(u_{j}^{(+n)},v_{j}^{(+n)}).
\end{eqnarray*}
Thus,
\begin{eqnarray*}
	f_{i}^{+},f_{i}^{-} & = & \sum_{n=0}^{N-1}\sum_{m=0}^{N-1}u_{n}^{(+0)}\left(u_{i}^{(-m)}u_{j}^{(+m)},v_{i}^{(-m)}u_{j}^{(+m)}\right)\\
	&  & +v_{n}^{(+0)}\left(u_{i}^{(-m)}v_{j}^{(+m)},v_{i}^{(-m)}v_{j}^{(+m)}\right)\\
	& = & \sum_{m=0}^{N-1}\left(u,v\right)_{i}^{(-m)}\sum_{m=0}^{N-1}u_{n}^{(+0)}u_{j}^{(+m)}+v_{n}^{(+0)}v_{j}^{(+m)}\\
	& = & \sum_{m=0}^{N-1}\left(u,v\right)_{i}^{(-m)}\left\{ d_{m}^{\dagger},d_{0}\right\} \\
	& = & u_{i}^{(-0)},v_{i}^{(-0)}.
\end{eqnarray*}
Additionally, this equivalence proves that
\begin{equation}
\left(u_{i}^{(-0)},v_{i}^{(-0)}\right)=\sum_{j=0}^{N-1}u_{j}^{(+0)}\left(F_{i,j}^{\prime},C_{i,j}\right)+v_{j}^{(+0)}(C_{i,j}^{\prime},F_{i,j})\label{eq:uvcorr_relation}
\end{equation}
Thus, expressions have been found for the couplings $f_{i}^{+},f_{i}^{-}$,
and thus for $f_{j}$.

\subsection*{Ground-state coupling for spins}

The BdG formalism can also be useful for spins by recognising that
$\sigma_{j}^{x}$ can be related to $a_{j}^{\dagger}$ and $a_{j}$
by the Jordan-Wigner transformation (JWT). That is,
\[
S_{j}^{x}=\mid\langle0\mid\sigma_{j}^{x}\mid1\rangle\mid^{2}=\mid\langle0\mid\prod_{n=0}^{j-1}\left(2a_{n}a_{n}^{\dagger}-1\right)\left(a_{j}^{\dagger}+a_{j}\right)\mid1\rangle\mid^{2}.
\]
For the first site of the chain, $j=0$, the translation is straightforward
\begin{eqnarray*}
	S_{0}^{x} & = & \mid\langle0\mid a_{0}^{\dagger}+a_{0}\mid1\rangle\mid^{2}\\
	& = & \mid A_{0}^{+}+A_{0}^{-}\mid^{2}\\
	& = & \mid u_{0}^{(-0)}+v_{0}^{(-0)}\mid^{2}
\end{eqnarray*}
where $A_{i}^{-}=\langle 0 \mid a_{i}\mid 1 \rangle$ and $A_{i}^{+}=\langle 0 \mid a_{i}^\dagger\mid1\rangle$. At the second site, $j=1$, the translation is
\begin{eqnarray*}
	S_{1}^{x} & = & \mid\langle0\mid\left(2a_{0}a_{0}^{\dagger}-1\right)\left(a_{1}^{\dagger}+a_{1}\right)\mid1\rangle\mid^{2}\\
	& = & \mid2\langle0\mid\left(a_{0}a_{0}^{\dagger}\right)\left(a_{1}^{\dagger}+a_{1}\right)\mid1\rangle-\left(A_{1}^{+}+A_{1}^{-}\right)\mid^{2}
\end{eqnarray*}
which is more complicated but still simple enough to analyse.

An analysis of this reveals that
\begin{widetext}
\begin{align}
S_{1}^{x} & =\left|2\sum_{m=0}^{N-1}u_{m}^{(+0)}\left[F_{0m}^{\prime}\left(F_{01}+C_{01}\right)-C_{0m}\left(F_{01}^{\prime}+C_{01}^{\prime}\right)+\left(C_{1m}+F_{1m}^{\prime}\right)\left(C_{00}^{\prime}-\frac{1}{2}\right)\right]\right.\nonumber \\
& +\left.2\sum_{m=0}^{N-1}v_{m}^{(+0)}\left[C_{0m}^{\prime}\left(F_{01}+C_{01}\right)-F_{0m}\left(F_{01}^{\prime}+C_{01}^{\prime}\right)+\left(C_{1m}^{\prime}+F_{1m}\right)\left(C_{00}^{\prime}-\frac{1}{2}\right)\right]\right|^{2}.\label{eq:s1_corr}
\end{align}
\end{widetext}
Now, analytical expressions for sites $j=0,1$ with straightforward
interpretations have been found. Using eq. \ref{eq:uvcorr_relation}
multiple times, eq. \ref{eq:s1_corr} can be simplified to 
\begin{widetext}
\begin{eqnarray*}
	S_{1}^{x} & = & \left|2\left(F_{01}+C_{01}\right)u_{0}^{(-0)}-2\left(F_{01}^{\prime}+C_{01}^{\prime}\right)v_{0}^{(-0)}+2\left(C_{00}^{\prime}-\frac{1}{2}\right)\left(u_{1}^{(-0)}+v_{1}^{(-0)}\right)\right|^{2}\\
	& = & \sum_{i=0}^{1}w_{u,i}^{(-0)}u_{i}^{(-0)}+w_{v,i}^{(-0)}v_{i}^{(-0)}
\end{eqnarray*}
\end{widetext}
where 
\[
\begin{array}{ccccc}
&  & w_{u,0}^{(-0)} & = & 2\left(F_{01}+C_{01}\right)\\
&  & w_{v,0}^{(-0)} & = & -2\left(F_{01}^{\prime}+C_{01}^{\prime}\right)\\
w_{u,1}^{(-0)} & = & w_{v,1}^{(-0)} & = & 2C_{00}^{\prime}-1
\end{array}
\]
or, equivalently 
\[
\begin{array}{ccccc}
&  & w_{u,0}^{(-0)} & = & 2\left(F_{01}+C_{01}\right)\\
&  & w_{v,0}^{(-0)} & = & -2\left(F_{10}^{*}-C_{10}\right)\\
w_{u,1}^{(-0)} & = & w_{v,1}^{(-0)} & = & 1-2C_{00}
\end{array}.
\]

\section{Intrinsic edge signatures of the spin chain}\label{app:edge-sig}

If for simplicity we focus on the immediate neighbour of the edge
site, the two probes $\langle0\mid\sigma_{0}^{x}\mid1\rangle$ and
$\langle0\mid\sigma_{1}^{x}\mid1\rangle$ can be represented by the
JWT as $S_{0}^{x}=\langle0\mid(a_{0}+a_{0}^{\dagger})\mid1\rangle$
and $S_{1}^{x}=\langle0\mid(2a_{0}a_{0}^{\dagger}-1)(a_{1}+a_{1}^{\dagger})\mid1\rangle$,
respectively. The former can be represented, using the Bogoliubov
de-Gennes (BdG) formalism, as simply the combined strength of the
amplitudes of the zero mode $u_{0}^{(-0)}+v_{0}^{(-0)}$ at the edge
site, where $u_{i}^{(-n)},v_{i}^{(-n)}$ are the wave function amplitudes
for the $n$'th single-particle excitation state on site $i$. The
presence of the string operator of the JWT in the latter
means that $\sigma_{1}^{x}$ does not simply probe the strength of
the wave function of the MZM as is. Using BdG we can represent this
coupling as a weighted sum of the amplitudes $u_{i}^{(-0)},v_{i}^{(-0)}$
of the zero-mode on all the sites of the chain including the edge
\begin{equation}
S_{1}^{x}=\sum_{i=0}^{1}\left[w_{u,i}^{(-0)}u_{i}^{(-0)}+w_{v,i}^{(-0)}v_{i}^{(-0)}\right]
\end{equation}
where the weight factors can be expressed via the BCS correlation
functions $C_{i,j}=\langle0\mid a_{i}^{\dagger}a_{j}\mid0\rangle$,
$F_{i,j}=\langle0\mid a_{i}^{\dagger}a_{j}^{\dagger}\mid0\rangle$,
\begin{equation}
\begin{array}{ccccc}
&  & w_{u,0}^{(-0)} & = & 2(C_{0,1}+F_{0,1})\\
&  & w_{v,0}^{(-0)} & = & -2(F_{1,0}^{*}-C_{1,0})\\
w_{u,1}^{(-0)} & = & w_{v,1}^{(-0)} & = & 1-2C_{0,0}
\end{array}
\end{equation}
and hence the edge-bulk difference is expected to be much less pronounced.

\section{The asymmetric spin response to local spin perturbations in the XY-chain}\label{app:asym-response}

The asymmetric spin response can be understood in terms of the Kitaev
chain by recognising that $\sigma_{j}^{x}$ translates into the fictitious
perturbation $\left[\prod_{i=0}^{j-1}2a_{i}^{\dagger}a_{i}-1\right]\left(a_{j}^{\dagger}+\frac{\Delta}{\mid\Delta\mid}a_{j}\right)$
at the ideal points. For $\Delta>0$, this can be shown to be equivalent
to $\prod_{n=0}^{j-1}\left(-i\gamma_{n}\eta_{n}\right){\color{black}{\color{green}{\color{black}\gamma_{j}}}}=\gamma_{0}\prod_{n=1}^{j}\left(-i\eta_{n-1}\gamma_{n}\right)=\gamma_{0}\prod_{n=1}^{j}\left(2d_{n}^{\dagger}d_{n}-1\right)$.
Thus, $S_{j}^{x}=\mid\langle0\mid\gamma_{0}\left[\prod_{i=0}^{j-1}2d_{i}^{\dagger}d_{i}-1\right]\mid1\rangle\mid^{2}$.
At the ideal point, it is clear that $d_{i}^{\dagger}d_{i}\mid1\rangle=0$
for all $0<i<N$. Thus, $S_{j}^{x}=\mid\langle0\mid(-1)^{j}\gamma_{0}\mid1\rangle\mid^{2}=\mid\langle0\mid\gamma_{0}\mid1\rangle\mid^{2}$
at all $j$. This is equivalent to $S_{j}^{x}=\mid\langle0\mid\gamma_{0}{\color{black}d_{0}^{\dagger}}\mid0\rangle\mid^{2}=\frac{1}{4}\mid\langle0\mid\gamma_{0}\eta_{N-1}\mid0\rangle-i\langle0\mid\gamma_{0}\gamma_{0}\mid0\rangle\mid^{2}$
and thus $S_{j}^{x}=\frac{1}{4}\mid i\langle0\mid\left(2d_{0}^{\dagger}d_{0}-1\right)\mid0\rangle-i\langle0\mid\gamma_{0}\gamma_{0}\mid0\rangle\mid^{2}=1$.
If, however, the opposite point is taken, where $\Delta=-1$, it is
found that $\sigma_{j}^{x}=-i\left[\prod_{i=0}^{j-1}2a_{i}^{\dagger}a_{i}-1\right]\eta_{j}$.
By performing similar MF operator algebra, it can be shown that $S_{j}^{x}=|\langle0\mid\eta_{j-1}\eta_{j}\mid0\rangle|^{2}$.
These two MF operators are completely uncorrelated at the ideal points,
and thus $S_{j}^{x}=0$. This explains the asymmetry along $\Delta$
in terms of Majorana formalism. Away from these ideal points, the
values move away from $1$ and $0$ as correlations begin to break
or build up for positive or negative $\Delta$ respectively. However,
the boundary between these regions is sharply defined around the line
of $\Delta=0$, where the bulk gap closes in the Kitaev model. This
is further support for the features being related to the topological
order of the Kitaev chain.

\section{Results for a longer chains ($N>8$)}\label{app:longer-chains}

\begin{figure}[h]
	\begin{centering}
		\includegraphics[scale=0.40]{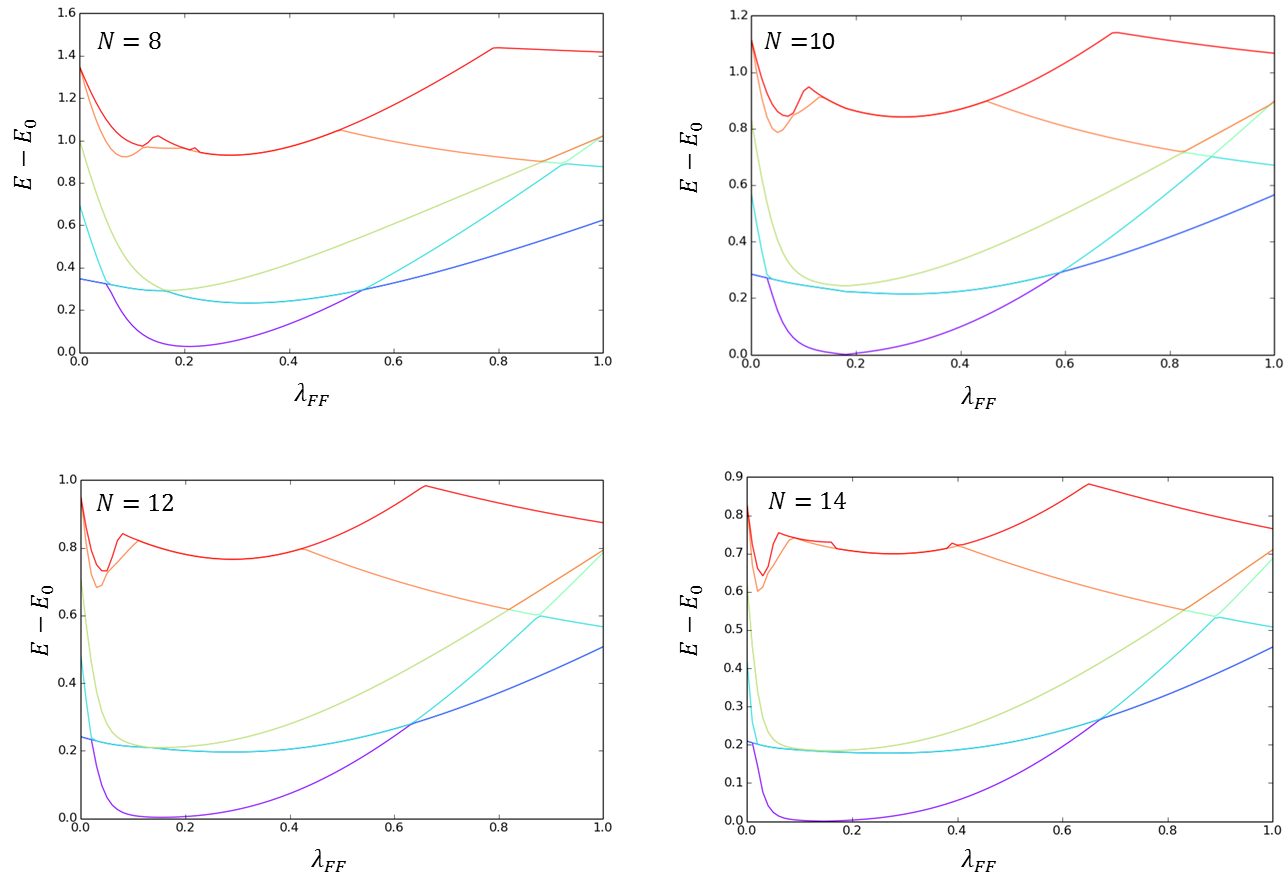}
     \par\end{centering}
	\caption{The transition towards protected doublet phase and increased degeneracy is shown as a function
		of the strength of non-local interactions $\lambda_{FF}$ for different chain lengths. The plots show energy differences ($E-E_0$) between the nth
		level denoted $E$ (plotted for levels $E=E_n$, n = 1; 2; 3; 4; 5) and the lowest energy
		level $E_0$ with parameters $\mu=0,t=1,\Delta=0$ as a function of the interaction strength $\lambda_{FF}$. The range of the parameter $\lambda_{FF}$ where strong degeneracy and protected phase exist is seen to increase with the length of the chain $N$.}
      \label{fig:lambda_range_N}
\end{figure}

One of the important features that was discussed is the appearance
of a gapped phase with a double quasi-degeneracy. This happens when
long range interactions of the 'flip-flop' type are added with sufficient
strength and the correct positive sign. One of the immediate questions
that is raised is whether the degeneracy is not absolute because of
a finite size effect. The trend can be clearly seen, \cref{fig:lambda_range_N}, by looking at
the spectra for increasingly large chains e.g. $N=8-14$ scanning
on the interaction parameter, $\lambda$, indicating that the phase is robust for larger chains.

\begin{figure}[h]
	\begin{centering}
		\includegraphics[scale=0.45]{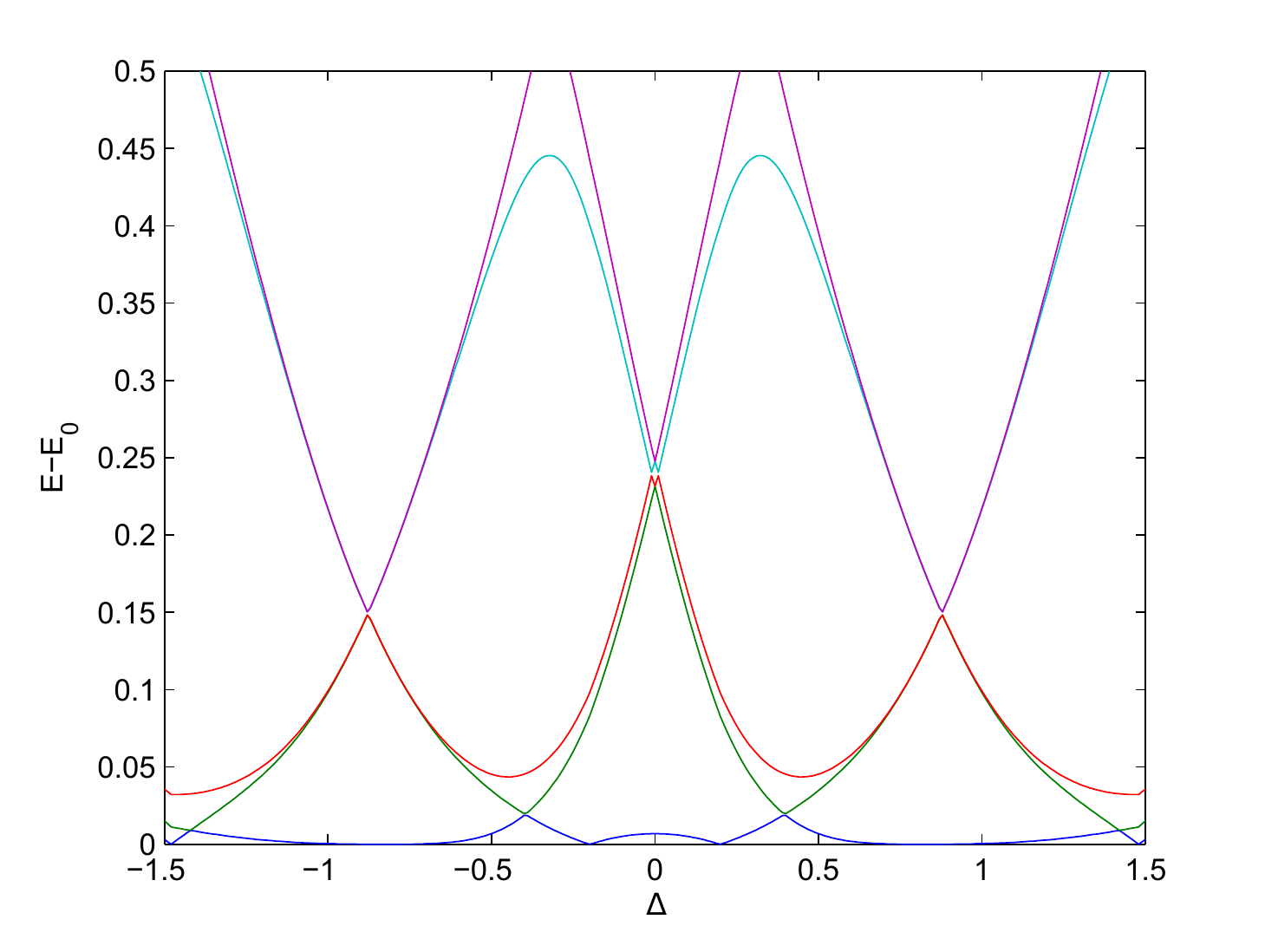}
	\end{centering}
	\caption{The transition towards the protected doublet phase shown as a function
		of the anisotropy parameter $\Delta$. The difference between the
		nth lowest energy level, E, (for n = 1; 2; 3; 4; 5) and the lowest
		energy level E0 for an N = 10 spin-chain with $\mu=0,t=1,\lambda_{FF}=0.15$
		.}
	\label{fig:higherN2}
\end{figure}

In \cref{fig:higherN2} the spectrum is shown as a function of the anisotropy parameter $\Delta$ and when compared to \cref{fig:specdiff} that even the incremental increase from $N=8$
to $N=10$ has a strong effect of strengthening the degeneracy. 

\section{Required extent of the interaction}\label{app:soft}

\begin{figure}[h]
	\begin{centering}
		\includegraphics[scale=0.4]{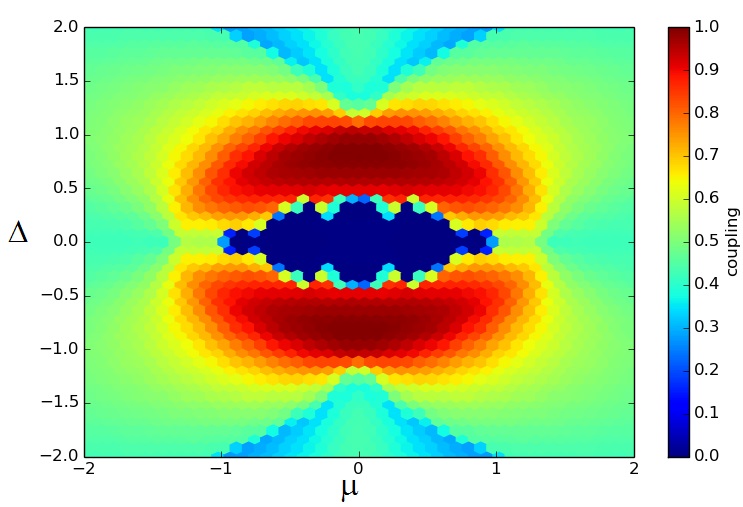}
		\par\end{centering}
	\caption{Using soft potential instead of a global all-to-all interaction indicates that while the range of the interaction needs to be long in order to find the protected states it can have an effective range and different functional forms. Here the potential is chosen to be exponential in the distance between sites $~\lambda\exp{(-\eta|i-j|)}$ with $\lambda=0.21$, an exponent of $\eta=0.1$ and chain length $N=8$.}
	\label{fig:expint}
\end{figure}

Realistic interaction potentials often have an effective range and a gradual decay and it is instructive to check the influence of interaction range on the extent of the protected states phase. For example, the results of choosing an exponentially decaying potential are shown in Fig.~\ref{fig:expint}. We observe that a significant extent of the interaction, measured in sites is required in order to observe the protected state but it does not need to cover the whole chain. We have checked other models of algebraic decay and we find similar results.

\section{Resilience against disorder}\label{app:dis}
In realistic physical systems we expect a certain variation in the local Hamiltonian parameters $\mu$, $t$ and $\lambda$. Since inside the protected phase there is a gap between the degenerate ground states and the first excited state we expect that small variations will not destroy the protection against external fields at least within some finite region.

\begin{figure}[h]
	\begin{centering}
		\includegraphics[scale=0.4]{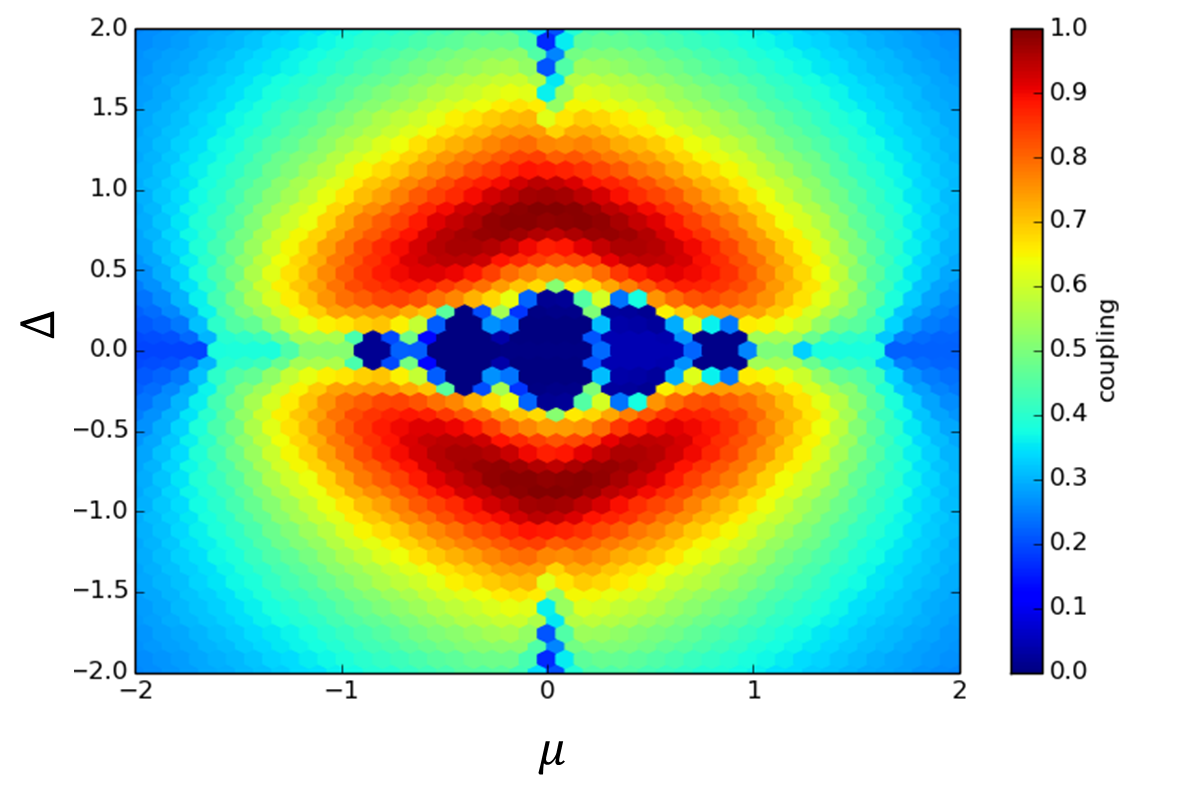}
		\includegraphics[scale=0.4]{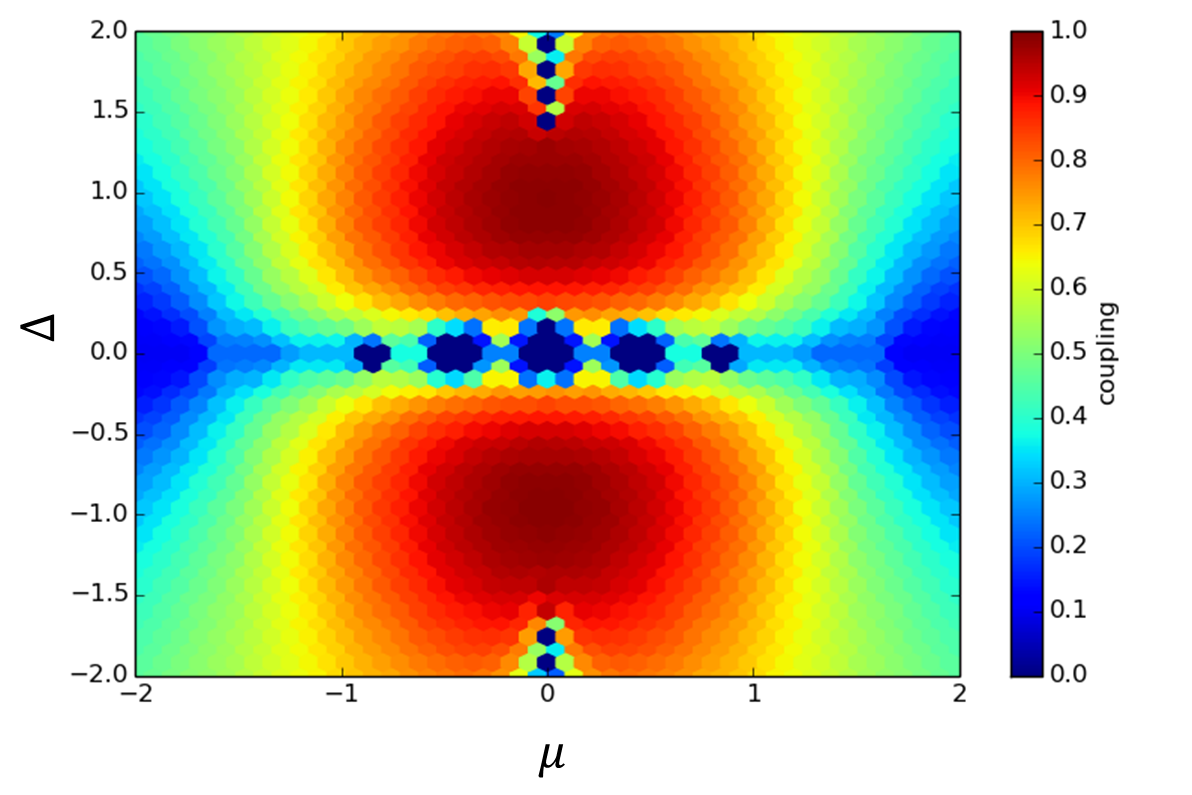}
		\par\end{centering}
	\caption{Examples of the influence of disorder in the parameters $\mu$ and $t$ on the region of protection in parameter space. In the upper panel a case with specific realisation of disorder ($10\%$ variation) in $\mu$ is shown and in the lower panel the disorder is in the coupling parameter $t$.}
	\label{fig:disorders}
\end{figure}
\begin{figure}[h]
	\begin{centering}
		\includegraphics[scale=0.4]{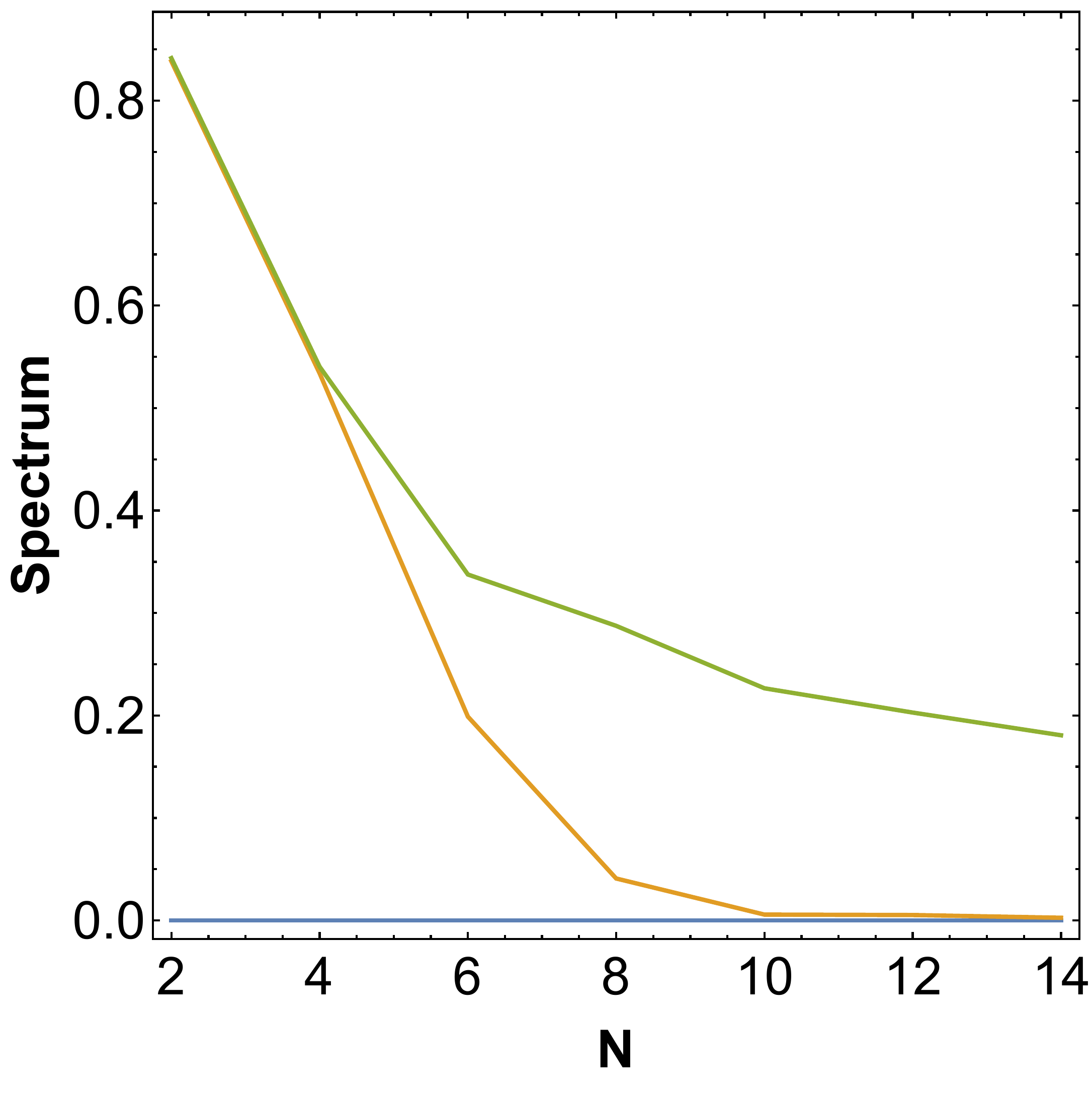}
		\par\end{centering}
	\caption{The energies of the lowest three states as a function of chain size $N$. Here the disorder is in the parameter $\mu$ with a strength of $5\%$ of the gap at $N=8$. The ground states are essentially remaining quasi-degenerate and the gap to the first excited state is maintained showing some finite tolerance to imperfections.}
	\label{fig:disorders_spect}
\end{figure}

In \cref{fig:disorders,fig:disorders_spect} we show an example of this. In \cref{fig:disorders} a specific random realisation of the Hamiltonian parameters is chosen where either $\mu$ or $t$ is varied to an extent of $\pm 10\%$ and the protected region is plotted. We see that even in the presence of such significant disorder there are region of protection. In \cref{fig:disorders_spect} the resilience of the degeneracy is tested and since perturbations will lift it when these are of the order of the spectral gap a variation of $\pm 5\%$ of the gap of $N=8$ is chosen. It can be seen that the degeneracy and gap remain resilient at this finite level of disorder.

\bibliography{bib}
\end{document}